\documentclass[aps,reprint,numerical,twocolumn,groupedaddress]{revtex4-1}
\usepackage{graphicx}% Include figure files
\usepackage{dcolumn}% Align table columns on decimal point
\usepackage{bm}% bold math
\usepackage{amsmath}
\usepackage{float}
\newcommand{\etal}{\mbox{\emph{et al.}}}

\begin{document}

%\preprint{}

%%%%%%%%%%%%%%%%%%%%%%%%%%%%%%%%%%%%%%%%%%%%%%%%% TITLE %%%%%%%%%%%%%%%%%%%%%%%%%%%%%%%%%%%%%%%%%%%%%%%%%%%%%%%%%%%%%%%%%%%%%%%%%%%%%%%%%%%%
\title{Magneto-optical spectroscopy with polarization-modulated light}
%%%%%%%%%%%%%%%%%%%%%%%%%%%%%%%%%%%%%%%%%%%%%%%%%%%%%%%%%%%%%%%%%%%%%%%%%%%%%%%%%%%%%%%%%%%%%%%%%%%%%%%%%%%%%%%%%%%%%%%%%%%%%%%%%%%%%%%%%%%%
%
\author{E. Breschi}
\email[]{evelina.breschi@unifr.ch}
\author{Z. D. Gruij\'c}
\author{P. Knowles}
\author{A. Weis}
%\thanks{}
\affiliation{Department of Physics, University of Fribourg, 1700
  Fribourg, Switzerland}

\date{\today}
%
%%%%%%%%%%%%%%%%%%%%%%%%%%%%%%%%%%%%%%%%%%%%%%%%% ABSTRACT %%%%%%%%%%%%%%%%%%%%%%%%%%%%%%%%%%%%%%%%%%%%%%%%%%%%%%%%%%%%%%%%%%%%%%%%%%%%%%%%%%%
\begin{abstract}
%%%%%%%%%%%%%%%%%%%%%%%%%%%%%%%%%%%%%%%%%%%%%%%%%%%%%%%%%%%%%%%%%%%%%%%%%%%%%%%%%%%%%%%%%%%%%%%%%%%%%%%%%%%%%%%%%%%%%%%%%%%%%%%%%%%%%%%%%%%%%%
%
We present a combined theoretical and experimental study of magnetic
resonance transitions induced by polarization-modulated light in
cesium vapor exposed to a transverse magnetic field.
Signals are obtained by phase-sensitive analysis of the light power
traversing the vapor cell at six harmonics of the polarization
modulation frequency.
Resonances appear whenever the Larmor frequency matches an integer
multiple of the modulation frequency.
We have further investigated the modifications of the spectra when
varying the modulation duty cycle.
The resonance amplitudes of both in-phase and quadrature components
are well described in terms of the Fourier coefficients of the
modulation function.
The background-free signals generated by the polarization modulation
scheme have a high application potential in atomic magnetometry.
\end{abstract}

\pacs{32.30.Dx, 07.55.Ge, 07.55.Jg, 78.20 Ls}

%33.40.+f    Multiple resonances in atomic and molecular physics
%33.57.+c    Magneto-optical and electro-optical spectra and effects
%76.70.Hb    Optically detected magnetic resonance (ODMR)
%
\maketitle
%
%%%%%%%%%%%%%%%%%%%%%%%%%%%%%%%%%%%%%%%%%%%%%%%%% INTRO %%%%%%%%%%%%%%%%%%%%%%%%%%%%%%%%%%%%%%%%%%%%%%%%%%%%%%%%%%%%%%%%%%%%%%%%%%%%%%%%%%%%%%%
\section{Introduction}
\label{sec:Int}
%%%%%%%%%%%%%%%%%%%%%%%%%%%%%%%%%%%%%%%%%%%%%%%%%%%%%%%%%%%%%%%%%%%%%%%%%%%%%%%%%%%%%%%%%%%%%%%%%%%%%%%%%%%%%%%%%%%%%%%%%%%%%%%%%%%%%%%%%%%%%%%

Magneto-optical spectroscopy of atomic media is a powerful tool that
uses resonant optical interactions for detecting magnetic resonance in
atoms (a comprehensive review is given by Alexandrov,
\etal{}~\cite{Alexandrov:2005:DEN}).
Magnetic resonance transitions are conventionally excited by a
magnetic field oscillating at the atoms' Larmor frequency, but it has
been known since the seminal work by Bell and Bloom in
1961~\cite{Bell:1961:ODS} that an amplitude-modulated (AM) resonant
light beam can also induce magnetic resonance transitions.
More recently, the Bell-Bloom method, in combination with
phase-sensitive detection, has found a renewed interest
\cite{Gawlik:2006:NMO, Schultze:2012:CPI}, in particular for its
application in high sensitivity atomic
magnetometry~\cite{Budker:2007:OM}.
Alternatively, modulation of either the light's frequency or
polarization at the Larmor frequency induces magnetic resonance.
While frequency modulation spectroscopy has become a well established
method for high sensitivity magnetometry \cite{Budker:2002:NMO,
  Andreeva:2002:CSD}, little work has been devoted to polarization
modulation~\cite{Aleksandrov:1973:OMI, Gilles:1991:OPP,
  Klepel:1992:TOP,Fescenko:2013:BBE}.
Recently, resonant polarization modulation has been used to solve
specific technical issues.
The authors of \cite{Ben-Kish:2010:DZF} have demonstrated that
polarization modulation between circular and linear polarization
states eliminates both dead-zones and heading errors in an alkali atom
magnetometer.
Polarization modulation between circular polarization states at the
ground state hyperfine transition frequency was shown to increase the
contrast of the clock resonance as described in~\cite{Jau:2007:PPL}
and references therein.

A theoretical model deriving algebraic expressions for the rich
resonance structure of magnetic resonance transition driven by
amplitude, frequency, and polarization modulation has recently been
presented~\cite{Grujic:2013:AMR}.
So far the model predictions were shown to give an excellent
description of experimental results obtained with amplitude modulation
using phase-sensitive detection~\cite{Grujic:2013:AMR} and for
polarization modulation with low-pass-filtered
detection~\cite{Fescenko:2013:BBE}.
Here, we address polarization modulation in combination with
phase-sensitive detection.
We characterize, both experimentally and theoretically, the
magneto-optical resonance spectra observed in a transverse magnetic
field, when the helicity of the exciting laser beam is periodically
reversed with different duty cycles.
We first introduce our experimental and theoretical methods, and then
discuss the full magneto-optical resonance spectra as a function of
the modulation duty cycle.
We show that resonances appear when the Larmor frequency is a harmonic
of the modulation frequency, and that the relative resonance
amplitudes are well described in terms of the Fourier coefficients of
the polarization modulation function.

%%%%%%%%%%%%%%%%%%%%%%%%%%%%%%%%%%%%%%%%%%%%%%%%% METHOD %%%%%%%%%%%%%%%%%%%%%%%%%%%%%%%%%%%%%%%%%%%%%%%%%%%%%%%%%%%%%%%%%%%%%%%%%%%%%%%%%%%%%%
\section{Method}
\label{sec:Method} 
%%%%%%%%%%%%%%%%%%%%%%%%%%%%%%%%%%%%%%%%%%%%%%%%%%%%%%%%%%%%%%%%%%%%%%%%%%%%%%%%%%%%%%%%%%%%%%%%%%%%%%%%%%%%%%%%%%%%%%%%%%%%%%%%%%%%%%%%%%%%%%%

Experiments are carried out in a conventional magneto-optical
spectroscopy apparatus~\cite{Breschi:2013:SCM,Fescenko:2013:BBE}.
Light from a 894~nm distributed feedback diode laser passes an
electro-optical modulator which flips the light polarization between
$\sigma^+$ and $\sigma^-$ polarizations at a constant frequency
$\omega_{mod}= 2 \pi \times 267$~Hz.
The degree of circular polarization was determined to be $99.7\%$ and
$99.8\%$, for $\sigma^+$ and $\sigma^-$, respectively.
The laser frequency can be tuned to any hyperfine component of the
cesium D$_1$-line ($6S_{1/2} \rightarrow 6P_{1/2}$), however, the
experiments reported here were carried out on the $F_g{=}4 \rightarrow
F_e{=}3$ transition which provides the highest magneto-optical
resonance contrast~\cite{Auzinsh:2008:RMO}.
The modulated beam passes an evacuated paraffin--wall-coated spherical
cell containing cesium vapor at room temperature ($20^{\circ}$ C)
saturated vapor pressure.
The laser beam diameter is about $10$ times smaller than the cell
diameter ($d=30$~mm) so that lensing or birefringence effects from the
cell itself do not critically affect the polarization quality.
This is testable in that linear polarization components will create
resonances at half of the Larmor frequency due to the symmetry and
evolution of any created spin alignment.
We did not see resonances in the experimental spectra arising from
spin-alignment.
The final excellent agreement between model and experimental results
(c.f.~\S\ref{sec:Results}) strongly implies that the cell has
negligible effect on the light polarization.

The cell is isolated in a magnetically controlled
environment~\cite{Breschi:2013:SCM}, with nominal residual field
components below a few nT\@.
The amplitude of a static magnetic field applied orthogonally to the
laser propagation direction, and hence the Larmor frequency
$\omega_L$, is scanned in the range $\pm4\,\omega_{mod}$.
The experimental signal is formed by detecting the light power $P(t)$
transmitted through the cell:
the photocurrent signal is converted into voltage, amplified, and then
demodulated via a lock-in amplifier which allows the simultaneous
recording of the in-phase and quadrature signals at six harmonics
$q\,\omega_{mod}$ of the modulation frequency.
%

%%%%%%%%%%%%%%%%%%%%%%%%%%%%%%%%%%%%%%%%%%%%%%%%%%%%%%%%%%%%%%%%%%%%%%%%%%%%%%%%%%%%%%%%%%%%%%%%%%%%%%%%%%%%%%%%%%%%%%%%%%%%%%%%%%%%%%%%%%%%%%%
\subsection{Model}
%%%%%%%%%%%%%%%%%%%%%%%%%%%%%%%%%%%%%%%%%%%%%%%%%%%%%%%%%%%%%%%%%%%%%%%%%%%%%%%%%%%%%%%%%%%%%%%%%%%%%%%%%%%%%%%%%%%%%%%%%%%%%%%%%%%%%%%%%%%%%%%
%
For signal interpretation we apply the model developed
in~\cite{Grujic:2013:AMR} for a spin-oriented atomic sample under
periodically modulated excitation and detection.
The key results of that model are summarized here.
%
%
%The power transmitted through the cell by an optically thin atomic
%vapor of length $L$ is given by the Lambert--Beer law,
%$P=P_0-P_0\,\kappa L$, where $\kappa$ is the optical absorption
%coefficient, and $L$ the sample length.
%
The power transmitted through the cell is given by the Lambert--Beer
law, which, for an optically thin atomic vapor of length $L$, is well
approximated by $P=P_0-P_0\,\kappa L$, where $\kappa$ is the optical
absorption coefficient, and $L$ the sample length.
For a laser beam resonant with a $F_g \rightarrow F_e$ hyperfine
transition, incident on atoms with a longitudinal spin orientation
$S_z$, the absorption coefficient becomes $\kappa = \kappa_0
(1-\alpha\,\xi\,S_z)$.
Here, $\kappa_0$ is the peak absorption coefficient of unpolarized
atoms, $\xi$, the light helicity, and $\alpha\equiv\alpha_{F_g,F_e}$
the orientation analyzing power.
Dropping time independent terms, the transmitted laser power is given by
\begin{align}
    P(t) &=  (\alpha \kappa_0 L P_0)\, S_z(t) \xi(t)\,,
\label{eqn:P(t)}
\end{align}
where $S_z(t)$ is the time dependent longitudinal spin orientation
resulting from the time-dependent solutions of the Bloch equations
under polarization modulation.  A square-wave modulation function
$\xi(t)$ is chosen to make a fast switch between the two circular
polarization states without spending significant time in
linearly-polarized intermediate states.
We allow for modulation with arbitrary duty cycle ($0<\eta<1$) chosen
to be symmetric with respect to $t=0$ and constrained to have
$|\xi(t)|=1$, whose Fourier expansion is given by
\begin{align}
\xi(t) & =  \sum_{j=-\infty}^{+\infty} g_j(\eta) \cos{(j \omega t)}\,,\\
%\end{align}
%
%\begin{align}
 g_0(\eta) & = 2 \eta - 1
\quad \text{and} \quad
g_{j{\neq}0}(\eta)  =\frac{2}{\pi} \frac{\sin{(\pi j \eta)}}{j} \, .
\label{eq:sigma}
\end{align}

In-phase $I_q$ and quadrature $Q_q$ parts of $P(t)$ extracted by
demodulation at $q\,\omega_{mod}$ can be written as
\begin{align}
 I_q(\eta) = \sum_{m=-\infty}^{\infty} a_{q,m}(\eta) \mathcal{A}_m
\text{,} \quad
 Q_q(\eta) = \sum_{m=-\infty}^{\infty} d_{q,m}(\eta) \mathcal{D}_m \,,
\label{eq:datamodel}
\end{align}
where $\mathcal{A}_m$ and $\mathcal{D}_m$ are, respectively,
absorptive and dispersive Lorentzian resonances given by
\begin{align}
 \mathcal{A}_m &= \frac{\gamma^2}
             {(m\,\omega_{mod} - \omega_{L})^2+\gamma^2} \, ,
\label{eq:lorentzianA}\\
 \mathcal{D}_m &= \frac{\gamma(m\,\omega_{mod} - \omega_{L})}
             {(m\,\omega_{mod} - \omega_{L})^2+\gamma^2} \, ,
\label{eq:lorentzianD}
\end{align}
with amplitudes
\begin{align}
a_{q,m}(\eta) &= \sqrt{2} \alpha
                   \kappa_0 L \frac{P_0^2}{P_s}\,g_m(\eta)\left[g_{q-m}(\eta) + g_{q+m}(\eta)\right] \,,
\label{eq:ampA}\\
d_{q,m}(\eta) &= \sqrt{2} \alpha
                   \kappa_0 L \frac{P_0^2}{P_s}\,g_m(\eta)\left[g_{q-m}(\eta) - g_{q+m}(\eta)\right] \,,
\label{eq:ampD}
\end{align}
where $P_s$ is the optical pumping saturation power.
We note that the model was developed for the low power limit
($a_{q,m}$,$d_{q,m}\propto P_0^2$), and that the experiments were
carried out in that range.

%%%%%%%%%%%%%%%%%%%%%%%%%%%%%%%%%%%%%%%%%%%%%%%%% SPECTRA %%%%%%%%%%%%%%%%%%%%%%%%%%%%%%%%%%%%%%%%%%%%%%%%%%%%%%%%%%%%%%%%%%%%%%%%%%%%%%%%%%%%%
\section{Analysis of the resonance spectrum}
\label{sec:Results}
%%%%%%%%%%%%%%%%%%%%%%%%%%%%%%%%%%%%%%%%%%%%%%%%%%%%%%%%%%%%%%%%%%%%%%%%%%%%%%%%%%%%%%%%%%%%%%%%%%%%%%%%%%%%%%%%%%%%%%%%%%%%%%%%%%%%%%%%%%%%%%%

We have studied the demodulated signals as a function of the detection
harmonic $q$ and modulation duty-cycle $\eta$.
Since in the Fourier expansion of a symmetric $\eta=0.5$ square wave the even
coefficients vanish, one sees from Eqs.~\eqref{eq:ampA}
and~\eqref{eq:ampD} that lock-in resonances appear only
at odd $m$ and even $q$.
To compare experimental and theoretical results we normalize the
in-phase and quadrature signals to the highest amplitude absorptive
and dispersive signals, respectively, found when $m{=}1$, $q{=}2$ and
$\eta{=}0.5$.
No other scaling factors are needed.
The normalized model signals thus read
\begin{align}
    \frac{I_q(\eta)}{a_{2,1}} & =
             \sum_{m{=}-\infty}^{\infty} \frac{a_{q,m}}{a_{2,1}} \mathcal{A}_m \,,
\label{eq:Inphasenorm}
\end{align}
\begin{align}
\text{and} \quad
   \frac{Q_q(\eta)}{d_{2,1}}  & =
             \sum_{m{=}-\infty}^{\infty} \frac{d_{q,m}}{d_{2,1}} \mathcal{D}_m \,.
\label{eq:Quadraturenorm}
\end{align}
Theory also predicts $d_{2,1}/a_{2,1}=1/2$.

Experimental amplitudes are obtained by fitting the recorded data with
the absorptive and dispersive Lorentzians of
Eqs.~(\ref{eq:lorentzianA}) and~(\ref{eq:lorentzianD}), yielding fit
amplitudes that represent the model amplitudes \eqref{eq:ampA} and
\eqref{eq:ampD}, respectively.
We then normalize all fitted in-phase (quadrature) amplitudes to the in-phase
(quadrature) amplitude of the ($m=1$, $q=2$, $\eta=0.5$) resonance.
The experimental $d_{2,1}/a_{2,1}$ ratio was found to be $0.492(1)$,
in accord with the theory prediction.
In this way, we eliminate all secondary experimental parameters
affecting the signal, e.g., detector quantum efficiency,
current-voltage converter gains, etc.  The normalization of the model
predictions, of course, removes unknown theory parameters (e.g.,
$P_s$, $\alpha_{F_g,F_e}$, $\kappa_0$, etc).

\begin{figure}[b]
  \centerline{
    \setlength{\unitlength}{1mm}
    \begin{picture}(40,1)(0,0)
      \put(20,-2.7){\makebox(0,0)[c]{\large\sffamily$\eta=\,$0.50}}
    \end{picture} }
\includegraphics[width=1\columnwidth]{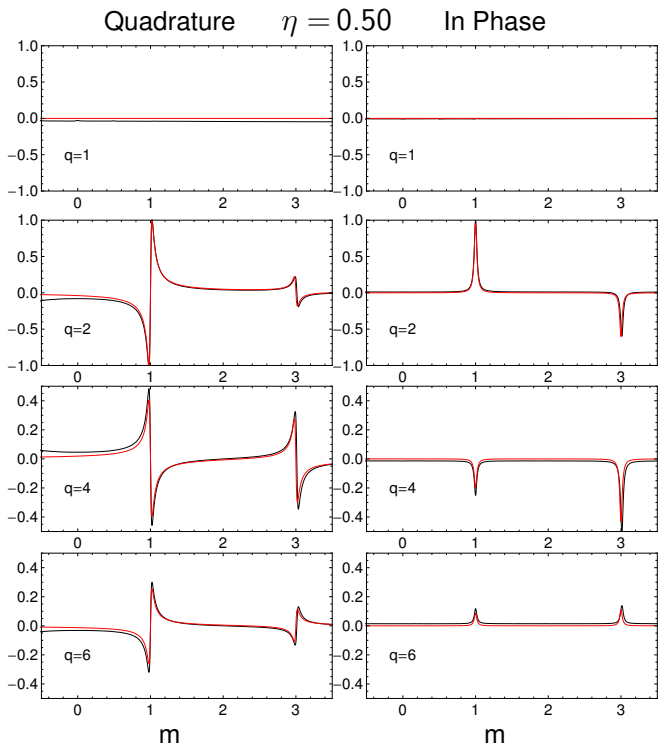}
 \caption{ (Color online) Magnetic field dependence at $50\%$ duty
   cycle ($\eta=0.50$) of normalized in-phase and quadrature lock-in
   signals, demodulated at even harmonics $q$ of $\omega_{mod}$; the
   odd harmonic signal is predicted to be zero, and the data (here
   $q=1$ is shown) supports the prediction.  Black lines are
   measurements, and red lines the model prediction. }
\label{fig:Fullspectrum05}
\end{figure}

\begin{figure}[b]
  \centerline{
    \setlength{\unitlength}{1mm}
    \begin{picture}(40,1)(0,0)
      \put(20,-2.7){\makebox(0,0)[c]{\large\sffamily$\eta=\,$0.10}}
    \end{picture} }
\includegraphics[width=1\columnwidth]{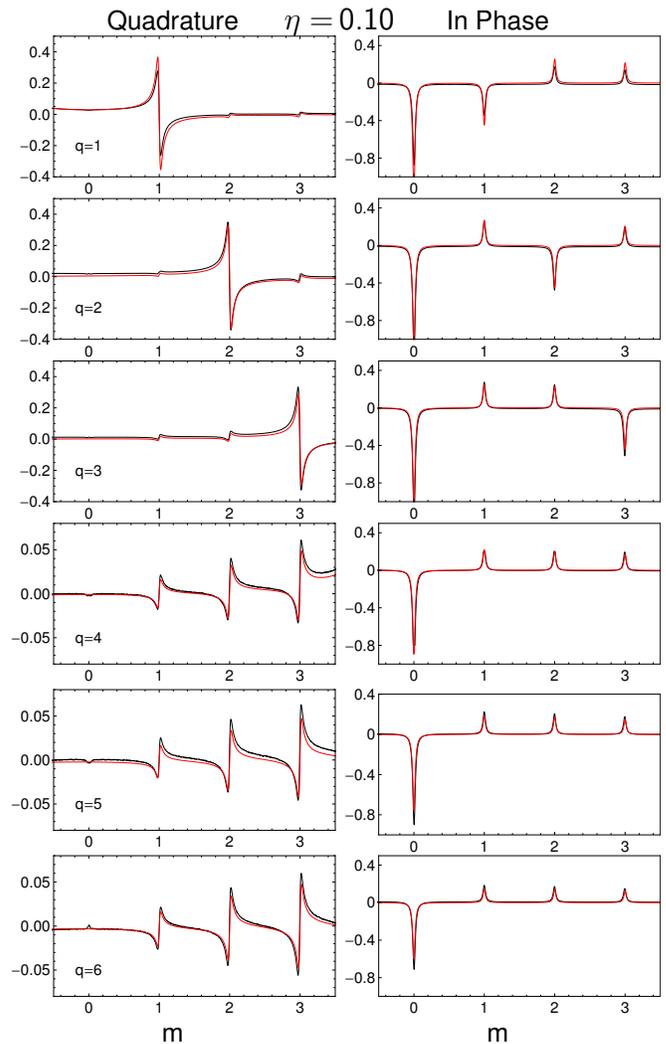}
 \caption{(Color online) Magnetic field dependence of normalized
   in-phase and quadrature lock-in signals, demodulated at harmonic
   $q$ of $\omega_{mod}$, for laser light resonant with the $F_g{=}4
   \rightarrow F_e{=}3$ transition.  The ratio between the Larmor
   frequency and the modulation frequency is $m$, with
   $m=\omega_L/\omega_{mod}$, where $\omega_{mod}= 2 \pi \times
   267$~Hz.  Black lines are measurements, and red lines the model
   prediction.  The duty cycle was $10\%$ ($\eta=0.10$). }
\label{fig:Fullspectrum}
\end{figure}

%\subsection{Duty-cycle}
In Fig.~\ref{fig:Fullspectrum05} and~\ref{fig:Fullspectrum}
we compare experimental and theoretical spectra for different values
of $q$ when $\eta=0.50$ and $\eta=0.10$.
No scaling has been applied beyond the 
normalization procedure above.

As predicted, resonances are observed when the Larmor frequency is a
multiple of the modulation frequency, i.e., $\omega_L= m
\omega_{mod}$.
An important point to note is that in contrast to experiments with
frequency- or amplitude-modulated light, polarization modulation
yields background-free in-phase and quadrature signals.
The amplitude of the zero-field level-crossing (Hanle) resonance is
proportional to $g_0$, Eq.~(\ref{eq:sigma}), and thus vanishes for
50\% duty cycle.

Figure~\ref{fig:Fullspectrum05} presents the signals obtained for
$\eta{=}0.50$, where one can see the largest signal at $m{=}1$ and
$q{=}2$ (the absorptive and dispersive reference resonances for
normalization).
This resonance is the most interesting for atomic magnetometry, and
work is ongoing to fully characterize the signal and study the final
sensitivity limit of the method.
The $\eta{=}0.5$ duty cycle pumping, i.e., excitation with a symmetric
square wave, essentially corresponds to the so-called
push-pull-optical-pumping~\cite{Jau:2004:PPO}, whose effect is the
synchronization of the pumping light's polarization modulation with
the harmonic evolution of the atomic quantum state in the external
field.
Notable in the $\eta{=}0.5$ case is that all $q=$odd resonances are
predicted to be zero by the model (as discussed at the start of the
section), and the data (only one trace for $q=1$ is plotted in the
figure) shows no signal at the detection limit in those cases.
Figure~\ref{fig:Fullspectrum} presents the signals obtained for
$\eta{=}0.10$, where $q=$odd resonances are non-null.

The magnetic resonance linewidths show no dependence on $m$ and $q$,
consistent with the fact that the linewidth is determined by
relaxation mechanisms and power broadening common to all resonances.
In the anti-relaxation wall-coated cell, the primary relaxation
mechanisms arise from spin-exchange collisions and loss of the atoms
to the alkali reservoir via the entrance channel, followed in
importance by collisions between alkali atoms and the anti-relaxation
coating, and magnetic field inhomogeneities, none of which have an
expected dependence on $m$ or $q$.

\begin{figure*}
\includegraphics[width= .8\textwidth]{{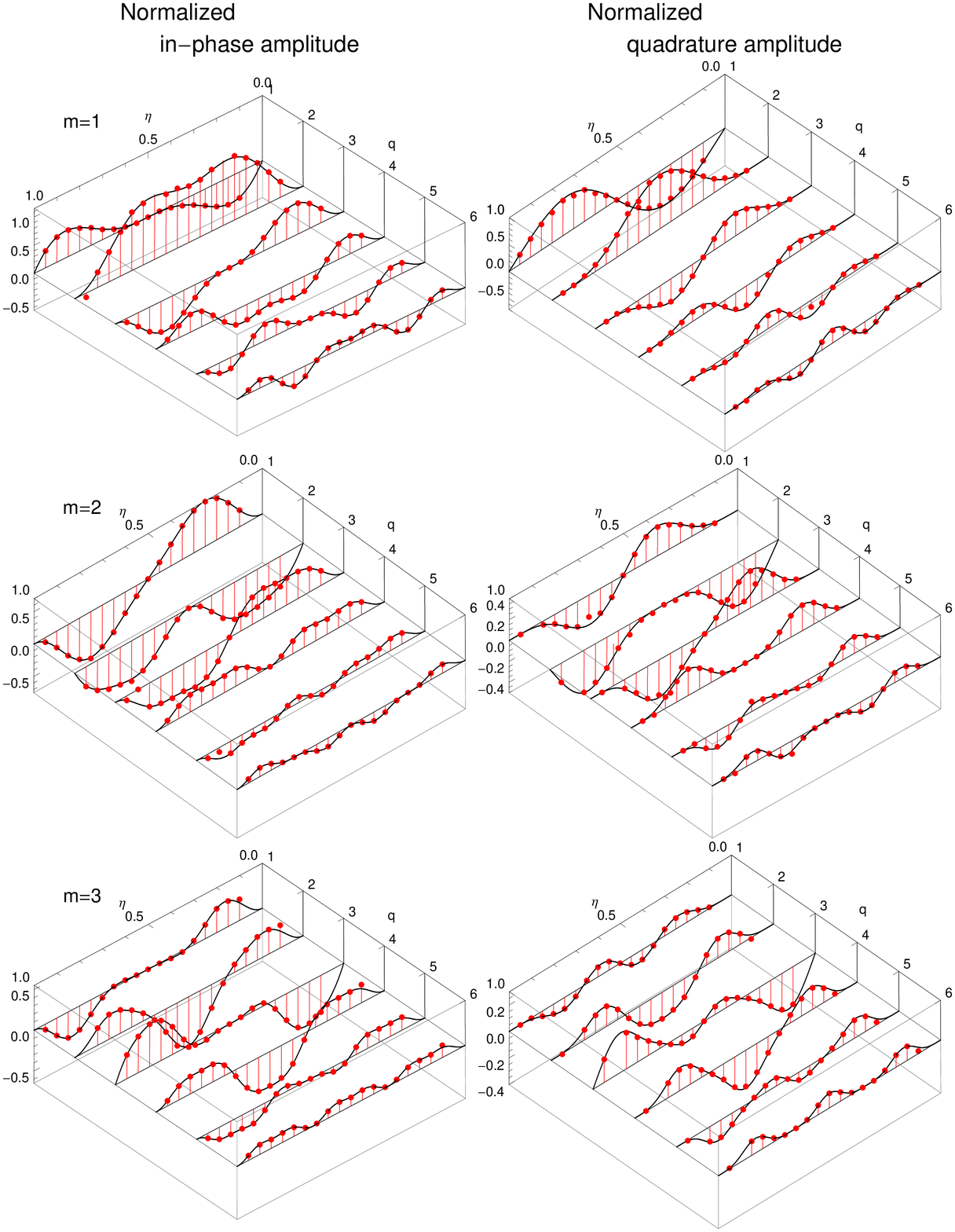}}
 \caption{(Color online) Normalized amplitude of the in-phase (left)
   and quadrature (right) signal for $m=1,2,3$ (from top to bottom
   line). The amplitudes are function of the detection harmonic
   ($q=1,\ldots,6$) and the duty cycle ($0< \eta <1$).}
\label{fig:transitionsAbs}
\end{figure*}

We have investigated the dependence of the spectra on the duty cycle
$\eta$ by varying $\eta$ between 0.1 and 0.9 in steps of 0.05.
For each value of $\eta$ we measured the amplitudes of the resonances
occurring at $\omega_L=\omega_{mod}$, $2\,\omega_{mod}$, and
$3\,\omega_{mod}$ (i.e., for $m=1,2,3$) in each of the six
demodulation channels $q=(1,\dots,6)\,\omega_{mod}$.
Figure~\ref{fig:transitionsAbs} shows the experimental amplitudes
(after normalization) as a function of $\eta$ and $q$, together with
the model predictions that are in excellent agreement with experiment.
We note the alternating symmetric/anti-symmetric $\eta$-dependence
with respect to $\eta=0.5$ for odd and even $q$, respectively.

The magnetic resonance linewidth depends on $\eta$ because the
effective power driving either the $\sigma^+$ or $\sigma^-$
transition--and hence the power-broadening of the line--depends on the
duty cycle $\eta$.
For this reason the balanced $\sigma^+/\sigma^-$ excitation with
$\eta=0.5$ is expected to produce the narrowest resonances, as
observed in experiment.

%%%%%%%%%%%%%%%%%%%%%%%%%%%%%%%%%%%%%%%%%%%%%%%% END  %%%%%%%%%%%%%%%%%%%%%%%%%%%%%%%%%%%%%%%%%%%%%%%%%%%%%%%%%%%%%%%%%%%%%%%%%%%%%%%%%%%%%%%%%
\section{Summary and Conclusions}
\label{sec:Con}
%%%%%%%%%%%%%%%%%%%%%%%%%%%%%%%%%%%%%%%%%%%%%%%%%%%%%%%%%%%%%%%%%%%%%%%%%%%%%%%%%%%%%%%%%%%%%%%%%%%%%%%%%%%%%%%%%%%%%%%%%%%%%%%%%%%%%%%%%%%%%%%

We have studied magneto-optical resonances that occur at multiples of
the Larmor frequency when the polarization of a resonant laser beam
traversing an alkali atom vapor is switched between left- and right-
circular polarization.
The lock-in demodulated signals have a rich spectral structure that is
well reproduced by algebraic model predictions.
The linear zero-crossings of the quadrature resonances can be used as
discriminator signals for magnetic field measurements.
The optimal resonance for magnetometry applications is the one with
$q=2, m=1$, and $\eta=0.5$, since it has the largest amplitude and
simultaneously the narrowest linewidth.
The application of the polarization modulation approach reported here
to atomic magnetometry has a distinctive advantage compared to similar
approaches using amplitude or frequency modulation, viz., the absence
of a DC background on the in-phase signal.
In a feedback-locked magnetometer, an imperfect phase setting
transfers a part of the DC background onto the quadrature signal
thereby introducing a shift in the magnetic field measure and
increasing power noise.

Finally polarization modulation with variable duty cycle can also find
applications in the preparation and manipulation of specific atomic
states, i.e., in metrology~\cite{DiDomenico:2010:CQS} and in quantum
information processing~\cite{Wang:2007:PDS}.

%
%%%%%%%%%%%%%%%%%%%%%%%%%%%%%%%%%%%%%%%%%%%%%%%%%%%%%%%%%%%%%%%%%%%%%%%%%%%%%%%%%%%%%%%%%%%%%%%%%%%%%%%%%%%%%%%%%%%%%%%%%%%%%%%%%%%%%%%%%%%%%%%
\appendix
%
%%%%%%%%%%%%%%%%%%%%%%%%%%%%%%%%%%%%%%%%%%%%%%%%%% THANKS %%%%%%%%%%%%%%%%%%%%%%%%%%%%%%%%%%%%%%%%%%%%%%%%%%%%%%%%%%%%%%%%%%%%%%%%%%%%%%%%%%%%%%
\begin{acknowledgments}
%%%%%%%%%%%%%%%%%%%%%%%%%%%%%%%%%%%%%%%%%%%%%%%%%%%%%%%%%%%%%%%%%%%%%%%%%%%%%%%%%%%%%%%%%%%%%%%%%%%%%%%%%%%%%%%%%%%%%%%%%%%%%%%%%%%%%%%%%%%%%%%%
%
This work is supported by SNF-Ambizione grant PZ00P2\_131926.
We thank the mechanical workshop and the electronics pool of the
Physics Department for expert technical support.
\end{acknowledgments}
%%%%%%%%%%%%%%%%%%%%%%%%%%%%%%%%%%%%%%%%%%%%%%%%%%%%%%%%%%%%%%%%%%%%%%%%%%%%%%%%%%%%%%%%%%%%%%%%%%%%%%%%%%%%%%%%%%%%%%%%%%%%%%%%%%%%%%%%%%%%%%%%
%
%%%%%%%%%%%%%%%%%%%%%%%%%%%%%%%%%%%%%%%%%%%%%%%%% BIBLIO %%%%%%%%%%%%%%%%%%%%%%%%%%%%%%%%%%%%%%%%%%%%%%%%%%%%%%%%%%%%%%%%%%%%%%%%%%%%%%%%%%%%%%
\raggedright
%% \bibliography{FRAPref}{}
%%
%% NB: warning, Breschi:2013:SCM in APB has not yet its volume
%% so it makes a weird extra comma.  This needs to be fixed in 
%% FRAPref when the volume becomes known finally.

%%%%%%%%%%%%%%%%%%%%%%%%%%%%%%%%%%%%%%%%%%%%%%%%%%%%%%%%%%%%%%%%%%%%%%%%%%%%%%%%%%%%%%%%%%%%%%%%%%%%%%%%%%%%%%%%%%%%%%%%%%%%%%%%%%%%%%%%%%%%%%%
%
\end{document}